\documentclass[sort&compress, times]{elsarticle}

\usepackage[latin1]{inputenc}
\usepackage[T1]{fontenc}
\usepackage[english]{babel}
\usepackage[babel=true]{csquotes}
\usepackage{enumitem}
\usepackage{layout}

\usepackage[top=3cm, bottom=3cm, left=2cm, right=2cm]{geometry}	% Page layout
\usepackage{graphicx}							%	Pour insérer des figures
\usepackage{wrapfig}							% Pour insérer les figures/images DANS le texte
\usepackage{caption}

\usepackage{amsmath}

\usepackage{lineno}

\usepackage{booktabs}
\usepackage{array}
\usepackage{tabularx}
\usepackage[flushleft]{threeparttable}		% Tableau en trois parties (simplifie les notes de bas de tableau)
\usepackage{multirow}
\usepackage{dcolumn}
\usepackage{rotating}		% Pour faire tourner les tableaux trop larges à 90° (mode paysage)

\usepackage{framed} % Framing content
\usepackage{multicol}

\usepackage[noprefix]{nomencl}																					% Nomenclature

\usepackage{hyperref}
\hypersetup{
    colorlinks = true
}

\makenomenclature

\newcommand{\degre}{^\circ}

\newcommand{\figref}[1]{\hyperref[#1]{\figurename~\ref{#1}}}
\newcommand{\tabref}[1]{\hyperref[#1]{\tablename~\ref{#1}}}

\newcommand{\etal}{\textit{et al.}}

\newcolumntype{k}{>{\global\let\currentrowstyle\relax}}
\newcolumntype{^}{>{\currentrowstyle}}
\newcommand{\rowstyle}[1]{\gdef\currentrowstyle{#1} #1\ignorespaces}
\newcolumntype{d}[1]{D{,}{,}{#1}}

\setlist{nosep}

\begin{document}

\begin{frontmatter}

\title{On the impact of the global energy policy framework on the development and sustainability of renewable power systems in Sub-Saharan Africa: the case of solar PV}

\author[unisinos]{Benjamin Pillot\corref{corresponding}}
\ead{benjaminfp@unisinos.br}

\author[corse]{Marc Muselli}
\ead{marc.muselli@univ-corse.fr}

\author[corse]{Philippe Poggi}
\ead{philippe.poggi@univ-corse.fr}

\author[unisinos]{Jo\~{a}o Batista Dias}
\ead{joaobd@unisinos.br}

\address[unisinos]{Universidade do Vale do Rio dos Sinos, Programa de P\'{o}s-Gradua\c c\~{a}o em Engenharia Mec\^{a}nica, Avenida Unisinos 950, 93022-000 S\~{a}o Leopoldo, Brasil}

\address[corse]{Universit\'{e} de Corse, UMR CNRS 6134 SPE, Route des Sanguinaires, 20000 Ajaccio, France}

\cortext[corresponding]{Corresponding author}

 %Abstract

\begin{abstract}

From 1972 and the \emph{Meadows} report to 1997 and the ratification of the Kyoto Protocol by most of the countries in the world, the global energy policy framework has undergone a paradigm shift, as human societies have been leaving aside the sole economic growth to embrace the broader concept of \emph{sustainable development}. As new electrification vectors, modern renewable power systems, such as solar PV, have sprung up in the wake of this major mutation and have typically been following an exponential growth since the mid-nineties. Electricity as well as its lack indeed holds a central position within the \emph{human development} apparatus, from highly interconnected centralized power networks reaching 100\,\% of the population in developed countries, to the almost 500 millions people without access to electricity in the multidimensionally poor rural areas of Sub-Saharan Africa. The encounter of the worldwide energy context being changed with each of these specific \emph{social-energy} paradigms has ended up considering two strategies for integrating renewable resources in power systems: the \emph{grid-connected} architecture for meeting the sustainable development requirements, and the rural \emph{off-grid} electrification for settling the lack of human development. By means of a historical and cross-sectional overview, we depict here the ins and outs of this protean pattern and, using the case of solar PV, analyze how it has eventually affected the development and sustainability of renewable power systems in Sub-Saharan Africa.

\end{abstract}


\end{frontmatter}

%\noindent Words:
%\begin{itemize}
%\item Words in text = 10 132
%\item Words in headers = 267
%\item Words outside text (captions, etc.) = 382
%\end{itemize}
%
%% Numérotation des lignes
%\linenumbers

\section{Introduction}

Energy and human development have been strongly correlated since the beginnings of time, and this assertion has become particularly true with the industrial era. The first industrialized countries in history have thus reached the highest level of human development and remain, at the present time, the largest consumers of energy per capita in the world \citep{pillot_2014, IEA_energy_balance_OCDE_2014}. Unfortunately, this evolution of the primary energy consumption, especially for sustaining the economic growth, which can be regarded as the \emph{quantitative} part of the human development \citep{HDR_2011, IEA_co2_emissions_2014}, has led to 2 main worldwide issues, that is exhaustion of the fossil fuel resources and global warming resulting from greenhouse gas (GHG) emissions.

The first oil shock in the seventies has raised the awareness of the developed countries about both these issues, outlined by the 1972 Meadows report about \emph{The limits to growth} \citep{meadows_1972}. The need of alternatives to petroleum, of a more efficient energy sector, of long-term energy policies or of an environmental approach to energy has become manifest to them \citep{IEA_history}. As a result, they have elaborated an international policy framework in order to meet these new goals, with at first the establishment of the International Energy Agency (IEA) in 1974 currently composed of 28 developed countries\footnote{According to the United Nations definition (see section \ref{sec: human development})} plus Turkey \citep{IEA_history, site_internet_iea}. This policy framework has then evolved with the increasing of the realization about the significant impact of economic development, along several key dates such as: 1983 and the designation of a World Commission about environment and development; 1987 and the Brundtland report defining the \emph{sustainable development} concept \citep{brundtland_1987}; 1992 and the \emph{Earth Summit} mainly resulting in a convention on climate change \citep{unfccc_1992}; 1997 and the Kyoto Protocol, the first international treaty inciting the industrialized countries to fight global warming by reducing their GHG emissions \citep{kyoto_protocol}. Accordingly, in order to overcome the world energy issue coupling fuel resource exhaustion and global warming, this specific policy framework has directly and indirectly been supporting the use of renewable energies for more than 40 years, but more significantly since the Kyoto Protocol \citep{kyoto_protocol, iea_renewable_energy_policies}.

Among all types of energy, electricity has more particularly defined modern life as we know it today, and is a very good marker of the level of development populations have reached \citep{hegedus_status_2003, mandelli_2016}. Many developing countries currently have really low electrification rates in comparison with the universal access to electricity of developed countries \citep{world_energy_outlook_2014, schafer_2011}. Within this context, rural areas of the developing countries are the most affected, especially in Sub-Saharan Africa\footnote{Here we consider Sub-Saharan Africa referring to all of Africa except the following countries: Algeria, Egypt, Libya, Morocco, Western Sahara, and Tunisia.}, where more than 80\,\% of the rural dwellers, that is 500 million people, are presently living without any electricity supply \citep{mainali_2013, mandelli_2016, world_energy_outlook_2014}. Accordingly, rural electrification has become a key parameter of the poverty reduction world policy \citep{mandelli_2016, zomers_2003}; among the different approaches proposed to meet this goal, the use of renewable power systems has been increasingly considered over the last 30 years \citep{schafer_2011, mandelli_2016}. However, in Sub-Saharan Africa, power supply capacity per capita has not increased since the eighties and the electrification rate still remains dramatically low \citep{HDR_2011, world_energy_outlook_2014}. On the other hand, many of the renewable systems currently used for off-grid rural electrification in developing countries also present sustainability and development issues \citep{nieuwenhout_2001, schafer_2011}.

At the present time, the solar potential is really substantial in Sub-Saharan Africa \citep{diabate_solar_2004, mohammed_2013}, which probably explains why, except traditional biomass, use of solar energy in rural areas is both more supported politically \citep{bugaje_renewable_2006, chineke_political_2010} and prevailing among field applications \citep{karekezi_pv_led_renewable_2002, ren_global_status_report_2015} than any other renewable resources. In the same way, photovoltaics (PV) is also the most promoted technology for electrifying rural people across the sub-continent, especially through solar home (SHS) and solar pico systems (SPS) \citep{karekezi_pv_led_renewable_2002, ren_global_status_report_2015}. However, the development of the sector is still in its beginning in Sub-Saharan Africa, with an effective installed capacity remaining low \citep{jager-waldau_pv_2014, ren_global_status_report_2015}, and, like other renewable power systems in developing countries \citep{schafer_2011}, field applications employed in off-grid electrification show the same sustainability issues \citep{karekezi_pv_led_renewable_2002, wamukonya_solar_2007, chaurey_2010}.

May this state of affairs partially be the consequence of the aforementioned world energy policy paragon set up by the developed countries, from the first appearance of the sustainable development concept to the current renewable energy policies derived from the Kyoto Protocol targets ? The thesis defended by the authors is indeed that the last 30 years of development in the renewable energy field have mainly been driven by one model, as a result of the global energy frame commitments coupled with the specific power supply structure of the developed countries, and that eventually this latter has negatively influenced the success of implementing renewable power systems in Sub-Saharan Africa. In order to understand the essence of this framework and how it has potentially impacted both the development and sustainability of the renewable energy projects in the region, we have achieved:
\begin{itemize}
\item an historical and cross-sectional overview of the global energy context, that is a comparative analysis of the ins and outs of the 2 different \emph{social-energy}\footnote{The \emph{social-energy} term features here the two-way interaction existing between energy and social frameworks, and emphasized in sections \ref{sec: section 3} and \ref{sec: section 4}.} paradigms involved, in order to apprehend the characterization of electricity supply and renewable power systems accordingly. From the need of \emph{sustainable development} to developing \emph{grid-connected} facilities on the one hand; from the lack of \emph{human development} to implementing rural \emph{off-grid} electrification on the other (section \ref{sec: section 2} to section \ref{sec: section 4});
\item a comprehensive description of the resulting impact of this pattern by focusing on the solar PV technology (section \ref{sec: section 5});
\end{itemize}

% \section{Global energy frame and renewable resources}
\section{Global energy policy framework}

\label{sec: section 2}

The current worldwide energy policy framework has a history whose the origin can probably be dated to the dawn of the seventies. It is rooted in the first oil shock and the related contemporary realization about the finiteness of the resources and the consequences of economic development, embodied by the \emph{Meadows} report\footnote{Though this report is probably the most well-known, it is important to cite the other founder, and more theoretical, work released by the economist Georgescu-Roegen one year before: \emph{The entropy law and the economic process} \citep{georgescu_1971}.}.

\subsection{Current energy situation}

Today, the global energy situation can be described as a crossing between two main issues: depletion of fossil resources and global warming due to greenhouse gas (GHG) emissions. The first one is related to the exhaustion of fossil fuels such as oil, coal, gas or uranium required to ensure operating of the classical energy supply systems. Reserves of these resources are indeed limited in time and decrease as fast as the world energy consumption increases. \tabref{tab: reserves of fossil fuels} depicts the current proved reserves of the main fossil fuels using the reserves to production ratio, which is an estimate of their lifetime according to the current energy consumption.

% Ratio consumption/production --> table ; graph evolution of GHG

\begin{table}[h]
\begin{center}

\begin{threeparttable}

\caption{Proved reserves of fossil fuels in 2013 \citep{iaea_2014, bp_2014}}

\label{tab: reserves of fossil fuels}

\begin{tabular}{@{\extracolsep{\fill}}cp{0.5cm}rl}
\toprule
Fossil fuel & \multicolumn{3}{c}{Reserves to Production ratio} \\
\midrule
Oil & & 53.3 & years  \\ 
Coal & & 111.2 & years \\ 
Natural gas & & 55.1 & years \\ 
Uranium\tnote{*} & & 128.3 & years \\
\bottomrule
\end{tabular}

\begin{tablenotes}

\footnotesize

\item[*] Reasonably assured resources + inferred resources

\normalsize

\end{tablenotes}

\end{threeparttable}

\end{center}
\end{table}

\begin{figure}[h]
\begin{center}
\includegraphics{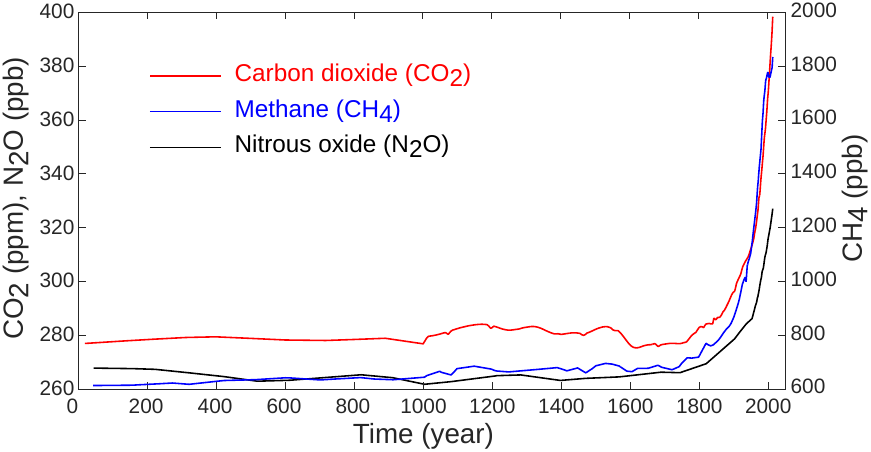}
\caption{Atmospheric concentration of the major anthropogenic greenhouse gases over the last 2000 years \citep{climate_change_2007_chapter2, epa_2015}, in parts per million (ppm) and parts per billion (ppb).}
\label{fig: ghg concentrations}
\end{center}
\end{figure}

This significant consumption of energy resources can be regarded as one of the main causes of the second issue, that is global warming. Except for nuclear energy, fuel combustion in the energy field is indeed the largest source of greenhouse gases (GHG), with almost 70\,\% of the global emissions, mainly consisting of carbon dioxide \citep{IEA_co2_emissions_2014}. This unnatural growth of the natural greenhouse effect has led to an increase of the average Earth's temperature \citep{climate_change_2007_chapter9}, and so to a sustainable change of the climate. Over the past 30 years, the world temperature has thus increased significantly \citep{climate_change_2007_chapter3}, while the atmospheric concentration of anthropogenic GHG (carbon dioxide, methane and nitrous oxide) has been growing dramatically, as depicted in \figref{fig: ghg concentrations}. At the present time, this relation between growth of the human industrial activity and global warming appears very consistent \citep{climate_change_2007_chapter9}, and so is almost unanimously admitted within the climate scientific community \citep{anderegg_expert_2010}.

It is within this specific questioning about human development issues that emerged, in the 1970s, new energy alternatives and the concept of \emph{sustainable development}.

\subsection{Advent of the sustainable development and rise of the renewable energies}

The first oil shock in 1973 was the start of a growing realization by the international community of the significance of the energy challenges in the mankind's future development. Beyond the Club of Rome's report, entitled \emph{The limits to growth} \citep{meadows_1972}, published in 1972, or the first international conference about environment which took place in Stockholm the same year \citep{liebard_amenagement_2005}, appears a more global questioning about the consequences of the lifestyle of the industrialized countries.

\subsubsection{Emergence of the \emph{sustainable development} concept}

The following step of this self-questioning was the designation in 1983, by the General Assembly of the United Nations (UN), of a World Commission on Environment and Development headed by Gro Harlem Brundtland. The report published by this commission in 1987, and entitled \emph{Our common future}, established in the first place the concept of \emph{sustainable development}, defined as a development \enquote{seeking to meet the needs and aspirations of the present without compromising the ability to meet those of the future} \citep{brundtland_1987}. Typically, this development is a balance between social, economic and environment features, so as to preserve the environment, the equity between peoples, persons and generations, and the economic efficiency \citep{liebard_amenagement_2005}. Thereafter, the Brundtland report was the base of the United Nations Conference on Environment and Development (UNCED), known as the \emph{Earth Summit}, which took place in Rio de Janeiro in 1992 \citep{liebard_amenagement_2005}. The United Nations Framework Convention on Climate Change (UNFCCC) adopted during this summit \citep{unfccc_1992} served then as a frame to the establishment and ratification of the Kyoto Protocol in 1997 \citep{kyoto_protocol}. The main purpose of this protocol, which has been ratified by 191 countries \citep{kyoto_protocol_web}, was to make 37 industrialized countries and the European Community (\emph{Annex I countries}) reduce their GHG emissions (Article 3), in order to promote sustainable development (Article 2). Accordingly, it has directly and indirectly encouraged these countries to look for alternative solutions in order to cut back their emissions, either by limiting their energy consumption or by replacing their current energy supply systems by lower emitter ones.

\subsubsection{Renewable energies}

Unlike fossil fuels, renewable energies are almost unlimited, at geological time scale, and carbon-free energy supply solutions. Indeed, these energies mainly come from the sun, directly or indirectly (direct solar, biomass, hydro, wind and wave energies), but also from the Earth's core high temperature (geothermal energy) and the Earth-Moon-Sun gravitational system (tide) \citep{ben_ahmed_consommation_2011, freris_renewable_2008}. \tabref{tab: yearly potential renewable} summarizes both the type and the yearly potential of each renewable resource. In addition, compared to the yearly world total primary energy supply ($155.5\times 10^{12}\,\mathrm{kWh}$), this table also shows that the annual renewable energy reserves are particularly significant, even if only a small part can really be exploited \citep{ben_ahmed_consommation_2011, IEA_energy_balance_non_OCDE_2014}.

\begin{table}[h]
\begin{center}

\caption{Estimated yearly potential of renewable resources \citep{ben_ahmed_consommation_2011}.}

\label{tab: yearly potential renewable}

\begin{tabular}{@{\extracolsep{\fill}}cccccc}
\toprule
 & \multirow{2}*{Sun} & \multirow{2}*{Biomass} & Water cycle & \multirow{2}*{Geothermal} & \multirow{2}*{Tide} \\
 & & & (hydro, wind, wave) & & \\
 \midrule
 Yearly & \multirow{2}*{$700\times10^{15}$} & \multirow{2}*{$4.5\times10^{15}$} & \multirow{2}*{$360\times10^{15}$} & \multirow{2}*{$300\times10^{12}$} & \multirow{2}*{$25\times10^{12}$} \\
potential (kWh) & & & & & \\
 \bottomrule
\end{tabular}

\end{center}
\end{table}

The second aspect differentiating renewable resources from fossil fuels is the amount of GHG emissions, especially of CO$_2$. In the case of the biomass, the fuel cycle is considered neutral, that is the CO$_2$ released during the resource combustion is counterbalanced by the CO$_2$ absorbed during its growth \citep{world_energy_outlook_2014}. For the other resources, they are either used directly or converted using systems which are not CO$_2$ emitters apart from the one indirectly released (manufacturing, transport, etc.). \tabref{tab: co2 emissions power plants} depicts the life cycle CO$_2$ emission rates of different power systems using fossil or renewable resources. Emissions of both these resources are significantly different, the emission rate being 10 to 20 times larger for fossil fuels than for renewable energies. Besides, the case of the nuclear energy remains particular since, despite low emissions, it presents other issues such as radioactive waste and installation safety.

% Tableau exemple japonais Hondo et al. : émissions GHG par type d'énergie

\begin{table}[h]
\begin{center}

\caption{Life cycle CO$_2$ emissions of fossil and renewable power generation, according to the Japanese example \citep{hondo_life_2005}.}
\label{tab: co2 emissions power plants}

\begin{tabular}{ccr}
\toprule
& Power generation & \multicolumn{1}{c}{CO$_2$ emissions} \\
& & \multicolumn{1}{c}{(gCO$_2$/kWh)}\\
\midrule
\multirow{4}*{Fossil} & Nuclear & 24.2 \\
\multirow{4}*{fuels} & Gas/combined cycle & 518.8 \\
& Gas/fired & 607.6 \\
& Oil & 742.1 \\
& Coal & 975.2 \\
\midrule
\multirow{3}*{Renewable} & Hydro & 11.3 \\
\multirow{3}*{systems} & Geothermal & 15.0 \\
& Wind & 29.5 \\
& Photovoltaics & 53.4 \\
\bottomrule
\end{tabular}

\end{center}
\end{table}

Hence, the renewable energies present inherent characteristics providing prospective solutions for the current world energy issues. Unlike fossil fuels, renewable resources are inexhaustible and converted through more ecological energy systems, and their reserves remain substantial enough for coping with the growth of the world population and so of the energy needs in the future.

\subsubsection{Sustainable development policy framework and new rise of the renewable energies}

As described previously, renewable resources present two main features directly related to the two issues of the global energy context, that is unlimited reserves and low GHG emissions allowing mitigation of both fossil fuel depletion and global warming effects. As a result, they have legitimately become one of the spearheads of the new main global energy policy.

The Kyoto Protocol has directly encouraged the Annex I countries to take into consideration renewable energies; indeed, the section 1.(a).(iv) of the article 2 recommends the \enquote{research on, and promotion, development and increased use of, new and renewable forms of energy} \citep{kyoto_protocol}. As a result, if all national renewable energy policies have been implemented after 1973, as a direct consequence of the first oil shock, about 95\,\% and 90\,\% of these policy measures have followed the establishment of the UNFCCC and the Protocol, respectively \citep{iea_renewable_energy_policies}. These incentive policies have essentially been relying on financial (public and private funding), fiscal (tax exemptions, environmental taxes), legislative (power purchase, grid access), political (penetration level targets, support programs) or technological and environmental support (R\&D program funding, Carbon Footprint concept) \citep{abdmouleh_2015}. In that way, even if hydro and geothermal energies were already used or considered at the end of the 19th century for producing electricity \citep{huacuz_photovoltaics_2003, secteur_electrique_2001, geothermal_energy_2015}, the emergence of the sustainable development, and thereafter of the corresponding energy policy framework (UNFCCC and Kyoto Protocol), has strongly fostered the new\footnote{Since mankind has firstly used renewable resources in its history (mills, sailboats, agriculture, etc.), we consider \emph{new} the use of these energies after the Industrial Revolution.} rise of the renewable energies.

\begin{figure}[h]
\begin{center}
\includegraphics{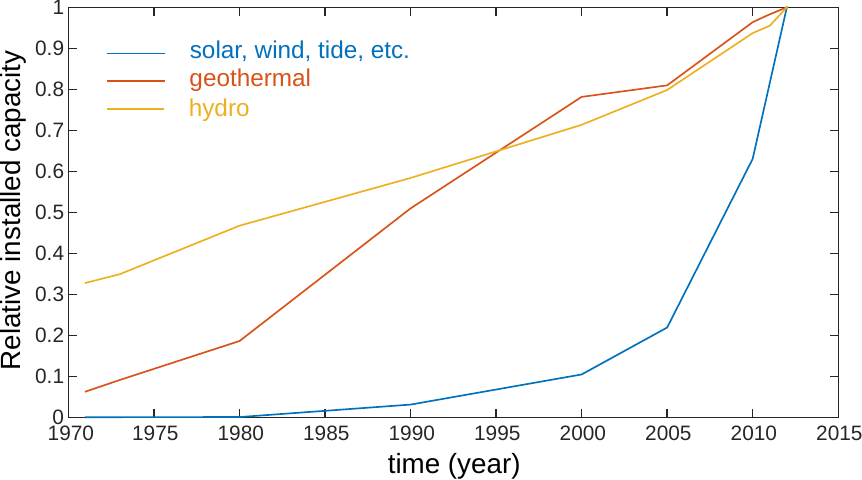}
\caption{Evolution of the renewable energy installed capacity over the past 40 years, for each technology \citep{IEA_energy_balance_non_OCDE_2014}.}
\label{fig: renewable evolution}
\end{center}
\end{figure}

Accordingly, since the 1970s, hydro energy generation has tripled, geothermal energy has been multiplied by 16 and production of energy from solar, wind, tide and other types has grown by almost a factor 2000 \citep{IEA_energy_balance_non_OCDE_2014}. Above all, the analysis of the evolution of these \emph{modern} renewable energies (solar, wind, tide, etc.), i.e. energies with no installed capacity before the 1970s, shows the significant role the energy policy framework has played in their development. \figref{fig: renewable evolution} depicts, for each renewable technology, the evolution of the relative installed power over the past 40 years, i.e. the ratio between the installed capacity at the time and the one corresponding to the current year (2012). If hydro and geothermal energies have approximately followed a linear increase, installed capacity of other renewable energies really sprung up in the 1980s and has had an exponential growth until now, with the incline getting significant at the beginning of the years 2000, that is after the establishment of the UNFCCC and the Kyoto Protocol.

\section{Electricity and development}

\label{sec: section 3}

Overall, development cannot be achieved without energy \citep{HDR_2011}. Total primary energy supply has thus grown more than twice in 40 years, mainly for supplying the economic growth \citep{IEA_co2_emissions_2014}. Within this specific paradigm, electricity mainly ensures enhancement of the qualitative criteria of the \emph{human development}.

\subsection{Human development and multidimensional poverty}

\label{sec: human development}

Today, the level of development of a country is defined by its human development index (HDI), ranging from 0 to 1, which is yearly calculated for more than 20 years by the United Nations Development Programme (UNDP) \citep{HDR_2014}. \emph{Human} development was originally defined by Mahbub ul Haq \citep{HDR_1990} as the growing process of both the choices available to the populations and the degree of well-being they have reached. In theory, these choices are infinite, but, in practice, some of them can be considered fundamentals since they allow access to all the other ones, that is having a long and healthy life, acquiring knowledge and getting enough resources for a decent standard of living \citep{HDR_1990}. The HDI is therefore computed from 3 indices assessing the level of quality of each of these criteria, that is health, education and income: the life expectancy index (based on the life expectancy at birth), the education index (based on the mean and expected time of schooling) and the gross national income (GNI) index (based on the GNI per capita).

The HDI was defined for 187 countries in 2013 \citep{HDR_2014}, divided into 4 quartiles corresponding to very high ($\mathrm{HDI} \geq 0.808$), high ($0.700\leq\mathrm{HDI}\leq 0.790$), medium ($0.556\leq\mathrm{HDI}\leq 0.698$) and low human development ($0.337\leq\mathrm{HDI}\leq 0.540$), respectively. Typically, the countries having a very high human development are considered \emph{developed} while all the other ones are regarded as \emph{developing}\footnote{The UN definition is different and considers the level of development in a geographical way, so that only Europe, Northern America, Japan and Australia/New Zealand are regarded as developed countries \citep{un_population_prospects_data_2014}.} \citep{HDR_2014}. Furthermore, although no official status exists for the most developed countries, one has nevertheless been attributed by the UN to 48 countries presenting the lowest HDI, known as the least developed countries (LDCs) \citep{tesfachew_2014}. With 33 out of 49 countries being part of the LDCs, Sub-Saharan Africa is by far the least developed region in the world \citep{HDR_2014}.

% Multidimensional poverty as a complement of the HDI

Moreover, for the least developed regions, it is also interesting to analyze the complement of the human development, that is the multidimensional poverty \citep{HDR_2010}, and its corresponding index, the multidimensional poverty index (MPI). If the HDI gives a global outlook of the level of development in a country, the MPI also indicates the part of the population suffering from deprivations and the intensity of these deprivations, in the same 3 dimensions (health, education and standard of living). Consistently, the MPI tends to increase when the HDI decreases, which means that human development must firstly rely on mitigation of the multidimensional poverty, especially when it affects most of the people such as in Sub-Saharan Africa, where more than 50\,\% of the population is currently multidimensionally poor  \citep{HDR_2010}. The 3 dimensions of the MPI are equally weighted, and are based on 10 indicators, corresponding to deprivation thresholds, distributed as follow \citep{HDR_2010}:
\begin{itemize}
\item 2 health thresholds related to having at least one household member who is malnourished and one or more children who have died;
\item 2 education thresholds related to having at least one school-age child not attending school and no household member who has completed 5 years of schooling;
\item 6 standard of living thresholds related to not having electricity, not having access to clean drinking water, not having access to adequate sanitation, using \enquote{dirty} cooking fuel (like wood or charcoal), having a home with a dirt floor, and owning a limited number of goods.
\end{itemize}

\subsection{Electricity in modern societies}

Among all energies, electricity is probably the most representative of the level of development of a country, and its advent at the end of the 19th century deeply changed the world behavior \citep{huacuz_photovoltaics_2003}. Energy supply per capita increased over the centuries with the technology intended to simplify the human daily activities; but electricity definitely allowed the rise of the \emph{modern era}.

As a matter of fact, electricity coupled with alternating current has provided a \emph{clean} and easy-to-use energy where it is consumed, and so has contributed to the current modern life. As a result, if electricity was about 2\,\% of the final energy in 1940, this share is now more than 18\,\% and steadily increasing \citep{ben_ahmed_consommation_2011, IEA_energy_balance_non_OCDE_2014}. Most of the devices (air conditioner, refrigerator, computer, etc.) and basic services (lighting, clean water, medical care, means of information and communication, etc.) making life more comfortable are now fueled by electricity \citep{huacuz_photovoltaics_2003}.

\subsection{Access to electricity: a rural issue}

If electricity allowed modernization of life in the rich countries, the access to that energy currently remains particularly disparate in the world, the electrification rate being strongly dependent on the level of development of each region.

In 2012, the electrification rate of the more developed regions was about 100.0\,\% against 76.3\,\% for the developing countries \citep{world_energy_outlook_2014}. Worldwide, it was almost 1.3 billion of people without any access to electricity. Also, in order to better understand the issue in developing countries, it is relevant to analyze the population distribution between urban and rural areas. Indeed, in 2012, urbanization rate of the more developed regions was 77.6\,\%, while it was equal to 47.3\,\% in developing countries \citep{un_population_prospects_data_2014}. Among them, only 29.9\,\% and 36.3\,\% of the people in LDCs and Sub-Saharan Africa, respectively, were living in urban areas. Essentially, more than 3 billion people are currently living in rural areas, and the linking of this social feature with the electrification rate highlights that, at the present time, about 83.6\,\% of the people without access to electricity are living in rural areas.

\begin{table}[h]
\begin{center}

\caption{Access to electricity in 2012 \citep{world_energy_outlook_2014, un_population_prospects_data_2014}.}

\small

\label{tab: electricity access}

\begin{threeparttable}

\begin{tabular}{@{\extracolsep{\fill}}kl^r^r^r^r}

\toprule

Region & Urbanization & Electrification & Urban electrification & Rural electrification \\
& rate (\%) & rate (\%) & rate (\%) & rate (\%) \\

\midrule
\rowstyle{\bfseries}More developed regions\tnote{1} & 77.6 & 100.0 & 100.0 & 100.0 \\
\hspace*{3mm} Japan & 91.9 & 100.0 & 100.0 & 100.0 \\
\hspace*{3mm} Australia/New Zealand & 88.6 & 100.0 & 100.0 & 100.0 \\
\hspace*{3mm} Northern America & 81.1 & 100.0 & 100.0 & 100.0 \\
\hspace*{3mm} Europe & 73.0 & 100.0 & 100.0 & 100.0 \\[5pt]

\rowstyle{\bfseries}Developing countries\tnote{1} & 47.3 & 76.3 & 91.1 & 64.0 \\
\hspace*{3mm} Latin America & 79.0 & 95.0 & 98.5 & 81.9 \\
\hspace*{3mm} Middle East & 69.0 & 91.7 & 98.3 & 78.0 \\
\hspace*{3mm} Developing Asia\tnote{2} & 43.5 & 83.0 & 95.2 & 74.4 \\
\hspace*{3mm} Africa & 39.2 & 42.5 & 68.0 & 25.6 \\
\rowstyle{\itshape}
\hspace*{6mm} Sub-Saharan Africa & 36.3 & 32.0 & 59.2 & 16.6 \\[5pt]

\rowstyle{\bfseries}World & 52.6 & 81.7 & 94.1 & 68.0 \\

\bottomrule

\end{tabular}

\begin{tablenotes}

\footnotesize

\item[1] According to the UN geographic definition.
\item[2] Corresponding to Asia without Middle East and Japan.

\normalsize

\end{tablenotes}

\end{threeparttable}
	
\end{center}
\end{table}

\tabref{tab: electricity access} summarizes this specific context, with global, rural and urban electrification rate, as well as the urbanization rate, for each region of the world. In developing countries, we can note that the access to electricity decreases while the proportion of people living in rural areas grows. Besides, with an electrification rate of 32\,\% and a power supply capacity per capita still identical to the 1980s \citep{HDR_2011}, Sub-Saharan Africa remains the least electrified region in the world. With 486 million people without access to electricity, the rural electrification rate of the region (16.6\,\%) is dramatically low, far behind the developing countries taken as a whole (64\,\%).

\subsection{Electrification as a vector of the human development}

\subsubsection{HDI vs access to electricity}

\begin{figure}[h!]
\begin{center}
\includegraphics{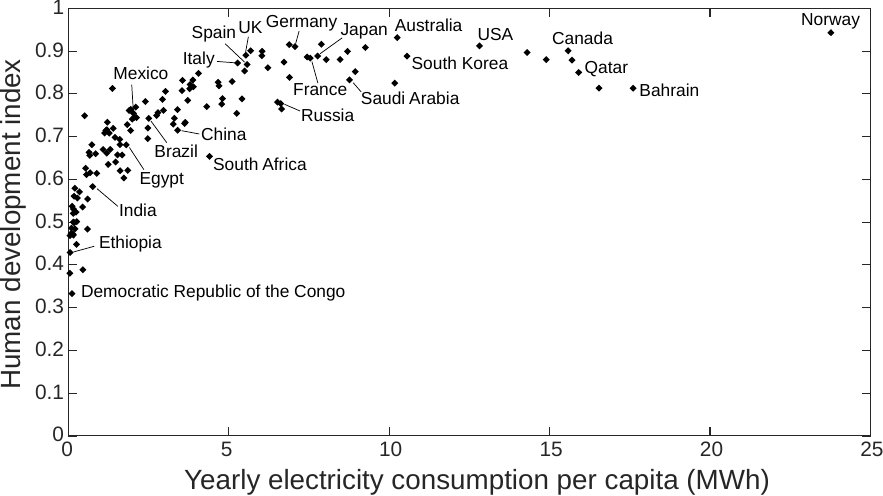}
\caption{HDI vs. electricity consumption \citep{HDR_2014, IEA_energy_balance_OCDE_2014, IEA_energy_balance_non_OCDE_2014}.}
\label{fig: hdi vs electricity consumption}
\end{center}
\end{figure}

\tabref{tab: electricity access} already gave an indication about the relationship between level of development and access to electricity. In fact, enhancement of the life conditions in many countries, through the rise of their HDI, would require increasing their electricity consumption by a factor 10 or more \citep{hegedus_status_2003}. Using values from 2012\footnote{Most recent available year for electricity consumption.} and according to the example of Hegedus and Luque \citep{hegedus_status_2003}, \figref{fig: hdi vs electricity consumption} illustrates the strong correlation existing between consumption of electricity and quality of life by plotting HDI versus electricity consumption per capita. As a result, it can be observed that the HDI asymptotically grows with electricity consumption, reaching its maximum threshold ($\mathrm{HDI}\approx0.9$) beyond approximately 5\,MWh/year.

\subsubsection{Effects of electricity on MPI components}

In order to understand how electricity access can enhance human development in the LDCs, and so in Sub-Saharan Africa, it is relevant to look at how it can remove the different deprivations in the 3 MPI dimensions. As a matter of fact, lack of electricity is more significant for people who are muldimensionally poor, and Sub-Saharan Africa remains the most affected region, with more than 60\,\% of the poor living without any access to it \citep{HDR_2010, HDR_2011}. Impact of a future electrification can be evaluated by analyzing the current issues the lack of electricity brings out, directly or indirectly assessed by the MPI, as well as the experiments conducted in different developing countries \citep{huacuz_photovoltaics_2003, HDR_2011, gustavsson_2007}.

By looking at every deprivation threshold, we can note that access to electricity is a critical criterion. It can contribute to the use of cleaner cooking fuels, especially for preventing diseases deriving from solid fuel combustion smoke \citep{baurzhan_2016}, to improve the access to clean water (pumping and filtering), to run some devices making life more comfortable. Regarding education, effects are either direct, such as the lack of individual or public lighting which restrains the available time for studying \citep{lahimer_2013, gustavsson_2007}; or indirect, with for example the time spent by children to collect water or cooking fuel which slows down academic progress and mitigates schooling rate, especially for the girls. In rural areas of Malawi, for example, they averagely dedicate about 4 hours per week for fetching water and wood, against 3 times less for the boys \citep{HDR_2011}. On the subject of health care, many modern devices and services (clean water, vaccine storage), which help to prevent child mortality, require electricity to operate, bad nutrition may partially result from inadequate cooking systems, the use of conventional energy resources and traditional light sources can decrease the air quality, and fuel fetching in difficult environments can be physically harmful \citep{HDR_2011, lahimer_2013}.

Initiatives against poverty in some developing countries also highlight the multidimensional positive influence of electricity on development, such as projects carried out by Rodrigo Baggio and Fabio Rosa in Brazil, or by Sanjit Bunker Roy in India. Baggio has democratized computer tools for young people in the favelas of Rio de Janeiro; Rosa has developed a solar panel rental company in Porto Alegre for rural people (fence electrification, irrigation, etc.); and Bunker Roy acts for education in rural areas of India, especially by using PV lighting to get schools up and running at night \citep{80_hommes}.

% Or
% to enable schools to function at night \citep{80_hommes}.
% especially by using PV lighting for schools to function at night
% especially by using PV lighting to get schools up and running at night

\subsubsection{Electricity and the qualitative part of human development}

\label{sec: electricity and qualitative part of human development}

As depicted previously, direct and indirect influences of electricity on the poverty dimensions are particularly significant. In addition, we can note that GHG emissions and primary energy supply are more correlated to the economic growth than to the evolution of the health and education criteria of the human development \citep{HDR_2011, IEA_co2_emissions_2014}. As a matter of fact, as shown by Klugman \citep{HDR_2011}, it is intuitively understandable that these fields first rely on a \emph{qualitative} use of energy, that is supplying energy for allowing services such as schools and hospitals to operate; on the contrary, the economic growth mainly relies on the manufacturing of goods and thus on a \emph{quantitative} use of energy. Considering the aforementioned relationship with reduction of multidimensional poverty, access to electricity therefore implies, in the first stages, enhancing the qualitative dimensions of the human development (health, education, life comfort), resulting \emph{de facto} in limited impact on GHG emissions and resource exhaustion.

\subsection{The rural issue in Sub-Saharan Africa}

Regarding power supply, if \tabref{tab: electricity access} confirms that the least developed regions in the world have the lowest electrification rates, it also highlights the significant disparities existing between rural and urban areas. Sub-Saharan Africa is an accurate picture of this issue since, with 486 million people, it is more than one third of the world population without access to electricity which is living in the rural areas of the region.

More than 60\,\% of the Sub-Saharan African population is currently living in rural areas, and projections show this prevalence is not likely to be reversed before 2040 \citep{un_population_prospects_2014}. Since human development is really low in most of the Sub-Saharan countries (33 are regarded as LDCs), rural populations are much more subject to poverty issues as they remain distant from the urban centers. As a matter of fact, in addition to global poverty, these populations also suffer from living in under-served remote areas \citep{hove_2013}, which is especially apparent concerning electricity network and transport infrastructures. The first one has already been emphasized with the low rural electrification rate (16.6\,\%); the second aspect is related to the very sparse road network, which consists of only 18\,\% paved roads, making the accessibility very low in the region, that is significant travel time to major cities \citep{nelson_2008}, and so any kind of supply (energy, material, etc.) is really tricky to achieve \citep{HDR_africa_2012}. Furthermore, infrastructural access to basic services, such as water supply or medical care, is also insufficient and more substantial in urban areas: access rates to improved sanitation facilities and improved water supply are respectively equal to 24\,\% and 40\,\% in rural areas, against 43\,\% and 82\,\% in urban areas \citep{HDR_africa_2012}. 

Inadequacy of electricity and transport networks makes the access to \emph{modern} fuels really difficult \citep{HDR_2011}, and incites people from rural areas to look for resources located in their close environment. As a result, in Sub-Saharan Africa, these populations mostly use fuelwood as energy supply \citep{bugaje_renewable_2006}, especially for cooking with nearly 730 million people (80\,\%) still relying on the traditional use of solid biomass \citep{world_energy_outlook_2014}. Depending on the context, this energy use alongside other kind of forestry and pastoral activities, such as cash crops, wood exportation or bushfires \citep{afrique_idees_recues_chapitre1}, can harm the surrounding environment significantly and therefore imply severe socioeconomic troubles for the populations \citep{bugaje_renewable_2006}. In this way, when the resulting local deforestation or desertification don't allow resources to be renewed anymore, these populations are then forced into rural flight \citep{HDR_2011}. Most often, they join the nearest urban centers, whose some of them like Timbuktu in Mali or Agadez in Niger have quadrupled their population in 30 years \citep{afrique_idees_recues_chapitre1}. As the polish author Ryszard Kapu\'{s}ci\'{n}ski once wrote \citep{ebene_kapuscinski}: \enquote{[these people] do not come here because the city needs them, but because poverty has driven them out from their village}. Finally, as urban centers also suffer from poverty and lack of infrastructures, they are not able to absorb this significant inflow, so these populations mostly come growing suburban slums and informal settlements, which makes human development still harder to enhance \citep{oyeyinka_2009, hove_2013}.

\begin{figure}[h]
\begin{center}
\includegraphics{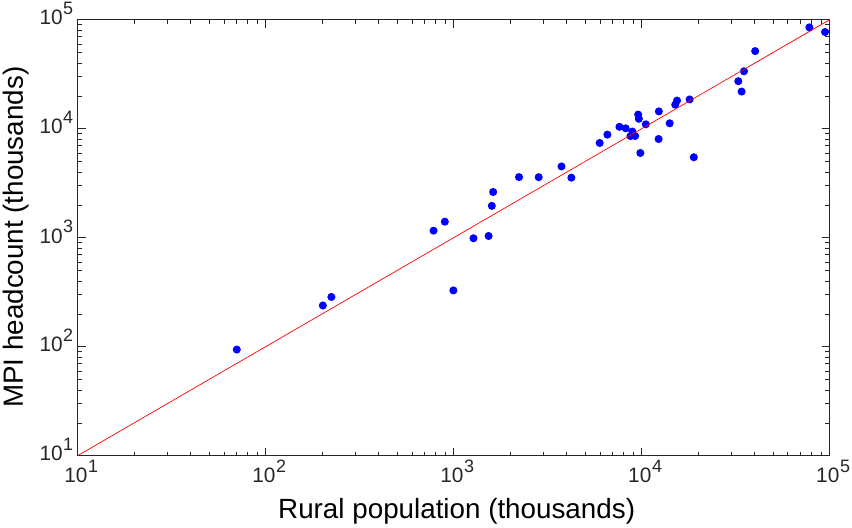}
\caption{Evolution of the MPI headcount with the rural population in Sub-Saharan countries \citep{HDR_2014, un_population_prospects_data_2014}.}
\label{fig: rural vs MPI}
\end{center}
\end{figure}

\figref{fig: rural vs MPI} summarizes this aggravated socioeconomic situation of rural people in Sub-Saharan Africa, by describing the increase of the number of people affected by multidimensional poverty (MPI headcount) with the rural population of a country. The almost linear evolution between both these criteria around the identity line indicates the strong likelihood for an individual living in rural areas of Sub-Saharan Africa to be multidimensionally poor.

% Former sentence :
% The almost linear evolution between both these criteria indicates the strong likelihood for an individual living in rural areas of Sub-Saharan Africa to be multidimensionally poor.

% "The almost linear evolution around the identity line"
% Almost linear evolution AND near the Identity line (1:1), since ONE more individual in rural areas means ONE more multidimensionally poor individual (y = 3x wouldn't really make sense...)

\section{Electrification paradigms and renewable energies}

\label{sec: section 4}

If energy has social implications, social structures also imply looking for suitable energy supply models; this influence of the one on the other allows us to speak of \emph{social-energy} issue. In the case of electricity, the rural context of Sub-Saharan Africa has led to a new major approach: the off-grid electrification.

\subsection{Centralized electrical networks}

Electricity was originally expensive, difficult to carry over long distances, and had to be directly consumed since storage was technically hard to achieve \citep{huacuz_photovoltaics_2003}. Alternating current (AC), by making electricity transmission technically and economically feasible, and interconnection, by ensuring the distribution of electricity without requiring any storage, have enabled to overcome these limits. By aggregating consumption and production hazards and using complementarity between power systems, interconnection has enhanced the reliability of the supply and reduced both the margins of safety and the fuel consumption of power plants \citep{secteur_electrique_2001}. Hence, the more powerful and interconnected a power station is, the more reliable and economical it will be (\emph{returns to scale} phenomenon); combined with AC use, it has then led to the actual \emph{centralized} structure of electrical grids \citep{secteur_electrique_2001, freris_renewable_2008}.

\begin{figure}[h]
\begin{center}
\includegraphics{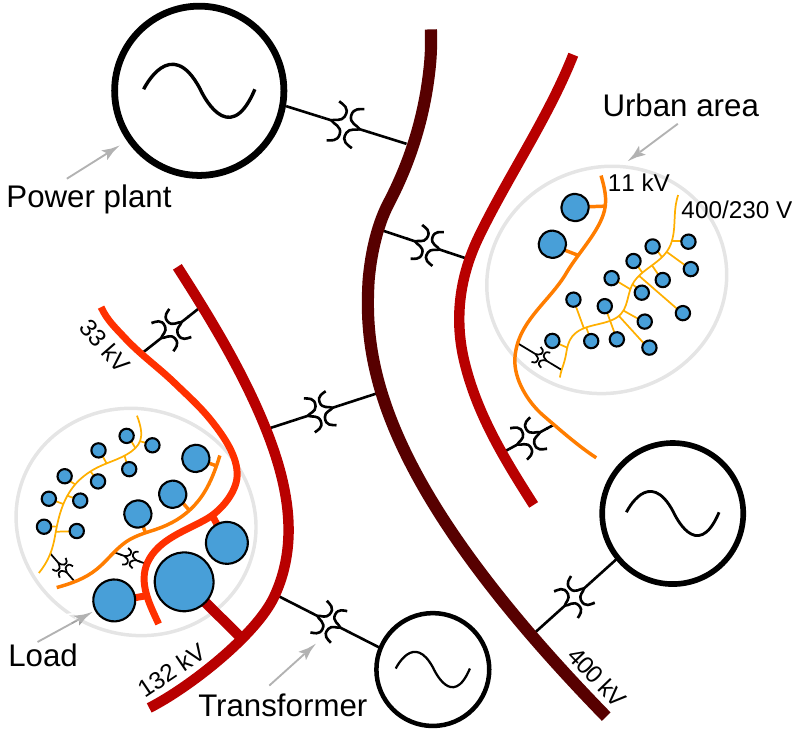}
\caption{Centralized grid, according to the UK example \citep{freris_renewable_2008}.}
\label{fig: centralized grid}
\end{center}
\end{figure}

A centralized grid is a large-scale power system carrying electricity from power stations to consumption sites (loads), through \emph{transmission} and \emph{distribution} networks, ensuring both transformation of electricity with respect to the consumer needs and permanent balance between production and consumption \citep{reseau_electrique_structure_1991}. Power plants are connected to the transmission network (extra high and high voltage), while energy is delivered to the consumers through the distribution network (medium and low voltage). \figref{fig: centralized grid} describes the structure of such a network using the UK voltage example \citep{freris_renewable_2008}. The transmission network (400 kV and 132 kV) carries electricity provided by power plants over long distances, while the distribution grid (33 kV, 11 kV and 400/230 V) supplies consumers in smaller areas.

\emph{Urban areas} stand here for the high urbanization rate of fully electrified regions combined with the electrification process history. In fact, in the beginnings, \emph{pseudo urbanization} followed the primary expansion of electrical grids, in a phenomenon known as \emph{pre-electrification}, as remote people was attracted to larger population centers \citep{huacuz_photovoltaics_2003}. This aggregation of the rural consumers allowed a critical load threshold to be reached, making the interconnected power system expansion in these areas economically relevant \citep{huacuz_photovoltaics_2003, deichmann_economics_2011}.

\subsection{Off-grid electrification and human development in Sub-Saharan rural areas}

In Sub-Saharan Africa, electricity supply can enhance human development of populations mainly located in rural areas, by mitigating multidimensional poverty, that is reducing deprivations and rural flight into urban slums. Electricity supply of these remote areas is currently almost non-existent, while urban electrification rate remains really low in the region, at 59.2\,\%, against more than 95\,\% in the other developing regions (\tabref{tab: electricity access}). Moreover, not only many Sub-Saharan countries have faced recurrent power crisis over the last years, pointing out the persistent decline of their centralized networks \citep{karekezi_renewables_meeting_poors_2002} and the lack of reliable power infrastructures \citep{moyo_2013}, but extending a network also requires proper load factors, and so enough gathering of population, to be economically viable \citep{deichmann_economics_2011, nassen_2002, nouni_2008}. As a result, these countries are likely more concerned at the present time with improving current grids and power infrastructures than about expanding it so as to reach small, remote and scattered rural areas, which appear on top of that economically less relevant than urban areas \citep{mandelli_2016}.

The current promoted idea for suiting the electricity needs of these populations can be defined, according to Huacuz and Gunaratne \citep{huacuz_photovoltaics_2003}, as the change of the electrification model from a centralized grid \emph{delivering electricity to rural areas} to a structure which allows \emph{increasing the access of the rural people to electricity}. The corresponding pattern is known as \emph{off-grid electrification}: providing electricity where it is consumed using small-scale and suited to the needs stand-alone power systems \citep{adkins_2010, pillot_2014, lahimer_2013, mandelli_2016, huacuz_photovoltaics_2003, pillot_ecm_2013}. Essentially, a centralized grid is a global power system mesh to which any consumer can connect, while off-grid electrification can be seen as a local customization of the power systems meeting the needs of every consumer.

\subsection{Renewable energies in power systems}

Depending on the electrification paradigm, the use and integration of renewable systems in the energy scheme is different. On the one hand, it means adapting the inherent intermittency and low capacity factors to the grid requirements, and on the other hand, meeting the local needs using small-scale and mixed power systems.

\subsubsection{Renewable energies in centralized power systems: grid-connected systems}

\label{sec: renewable grid-connected systems}

Because of their inherent features, integration of renewable energies in modern centralized grids presents some significant challenges. Firstly, unlike fossil fuel based power systems, renewable energies such as wind and solar present significant geographical scattering, and so often require small-scale conversion systems \citep{ben_ahmed_consommation_2011}. Therefore, most of the renewable power stations have to be directly connected onto the distribution network (\emph{decentralized production}) rather than onto the transmission system as usual \citep{freris_renewable_2008, kaundinya_2009}. Secondly, most of the renewable energies, including wind, solar, wave and tide energies, are \emph{intermittent}, because of climate features, meteorological phenomenons, geographical location or day-night alternation. This intermittency makes variable renewable energy systems have low capacity factors (variability) and causes temporal mismatch between demand and supply within the grid (uncertainty). As a result, integration of variable renewable energies require grid adaptations such as flexible generation, energy storage or geographical aggregation in order to maintain reliable and quality energy supply \citep{li_2015}. Even if grid-connected renewable systems using storage exist, they still remain confined to the research field or to some prototype projects for now \citep{subburaj_2015}. So, when speaking of grid-connected systems, they are tacitly regarded as systems directly connected to the grid without any kind of storage \citep{eltawil_2010, dowds_2015, kaundinya_2009}. For the time being, the intermittency issue and this lack of storage mainly lead to limit the penetration rate of renewable energies in electrical networks, especially in low and not interconnected power grids, such as island grids \citep{EDF_2010, etxegarai_2015, eltawil_2010}.

\subsubsection{Renewable energies in off-grid electrification: stand-alone systems}

\label{sec: renewable stand-alone systems}

Off-grid electrification using renewable energies is achieved with stand-alone power systems meeting the local demand of remote areas \citep{kaundinya_2009}. In that case, the installation almost always needs storage to suit demand and supply, and can rely on one or more renewable resources, include fossil fuel system (hybrid renewable system) and be designed as a mini-grid \citep{micro_reseaux_2009, kaundinya_2009, krishna_2015}. Furthermore, inherent characteristics of renewable energies are also more suited to the off-grid case. First, their geographical dispersion allows some of these resources to be almost always retrieved wherever remote people are located. Second, although low capacity factor and current storage technology limit available power ranges \citep{stockage_systemes_electriques_1998, kaundinya_2009}, it is of less concern in the case of rural electrification. Indeed, both population density, i.e. load factors, and electricity needs, particularly when the given rural area was un-electrified until then, are really low compared to urban centers \citep{huacuz_photovoltaics_2003, mandelli_2016}. Lastly, the small size of the corresponding conversion systems gives them some flexibility of use, and their modularity, such as in the case of PV, allows them to grow in pace with the load \citep{huacuz_photovoltaics_2003, kaundinya_2009}.

\section{Resulting issues of the rural off-grid electrification using renewable energies in Sub-Saharan Africa: the example of solar PV}

\label{sec: section 5}

Depending on the \emph{social-energy} paradigm, the use of renewable energies for providing electricity has been apprehended and promoted differently, either implemented in the original centralized electrification model or in the new off-grid pattern \citep{zomers_2003, mandelli_2016, secteur_electrique_2001, freris_renewable_2008}. The latter seems indeed the best approach for enhancing life conditions of the rural people in the least developed countries, especially in Sub-Saharan Africa where they stand for more than 60\,\% of the whole population . However, the success of this model shall rely on \emph{the capacity of the couple research-industry to develop concepts and systems suited to the Sub-Saharan context}. The current issues of the few off-grid renewable applications, such as PV systems, dispatched across the sub-continent demonstrate this is not necessarily the case \citep{nieuwenhout_2001, schafer_2011, karekezi_pv_led_renewable_2002, wamukonya_solar_2007, chaurey_2010}. We propose here to expose the impact the historical need for \emph{sustainable development}, combined with the specific power network structure of developed countries, has had on renewable systems development and sustainability in Sub-Saharan Africa by articulating our thesis around solar PV.

\subsection{Solar PV}

Photovoltaic cells encapsulated in flat modules are semi-conductors converting a part of the solar spectrum into electricity. The main component of any PV system is the PV electrical generator which is composed of several interconnected PV modules. At present time, the technology is mature and 2 main system configurations are sharing the market: the grid-connected and the stand-alone architectures.

\subsubsection{Grid-connected PV systems}

Overall, the grid-connected model consists in a PV array connected to the electrical network through a DC/AC inverter operating in phase with the grid \citep{energie_solaire_pv, eltawil_2010, obi_2016}. Electricity is directly injected into the power network and sold to the national electric utility through feed-in tariffs \citep{jager-waldau_pv_2014}. The grid-connected market is segmented between centralized, also known as \emph{utility-scale}, and decentralized power stations \citep{IEA_PVPS_2015, kaundinya_2009, preiser_photovoltaic_2003}. Typically, centralized grid-connected systems are large size PV power plants directly connected to the transmission grid, while decentralized systems are medium or small size applications, such as residential or commercial rooftop PV systems, connected to the distribution network \citep{obi_2016, preiser_photovoltaic_2003}. Also, the coupling of these systems with storage, mainly for smoothing the production in order to prevent grid instability issues due to intermittency, mostly remains at the prototype stage \citep{subburaj_2015}. The market is actually moving slowly because of the lack of markets where storage could be competitive, as a result of the cost of the storage solutions combined with the too few existing incentives \citep{IEA_PVPS_2015}. Like other renewable power sources, grid-connected PV systems are therefore mainly seen and implemented without any kind of storage \citep{eltawil_2010, obi_2016, energie_solaire_pv}.

\subsubsection{Stand-alone PV systems}

Stand-alone systems supply electricity to remote loads independently of the utility grid \citep{kaundinya_2009}, through 3 main configurations \citep{energie_solaire_pv, pillot_2014}:
\begin{description}
\item[Solar water pumping:] the most simple and economical systems, only consisting of a PV array connected to a pump, potentially through a controller/converter. Without any storage, the production is used at once and depends on the sunlight. This kind of system is mostly used for water pumping in hottest regions \citep{solar_pv_pumping_2008}.
\item[Classical stand-alone systems:] the most common systems, coupling PV modules with storage, such as solar home systems (SHS) \citep{wamukonya_solar_2007}. This configuration allows supply of the load without interruption, the storage device being charged during the day, taking over the PV generator during the night, and acting as an energy buffer in case. In order to maximize energy flows and to protect batteries from overload and deep discharge, these systems are most of the time equipped with a charge controller; inverters are also often implemented for addressing AC needs. Regarding storage, electrochemical batteries are by far the most used at present time, in particular the low-cost and well-known lead-acid technology \citep{uwe_sauer_storage_2003, al-sheikh_2012}; other means such as flywheels and fuel cells are also being studied \citep{darras_sizing_2010}.
\item[Hybrid stand-alone systems:] this configuration consists in integration of other energy sources within classical stand-alone systems, in order to enhance reliability and reduce investment costs in PV modules and storage. For the time being, fuel generators are the most common secondary energy source \citep{preiser_photovoltaic_2003}, but other models of hybridization exist such as wind/solar PV \citep{gergaud_modelisation_2002}.
\end{description}

\subsubsection{Solar PV: the spearhead of the renewable power systems in rural areas of Sub-Saharan Africa}

At the present time, modern renewable resources (hydro, solar, wind, geothermal and bioenergy) account for less than 2\,\% of the energy mix in Sub-Saharan Africa \citep{africa_energy_outlook_2014}. Among these energies, solar potential is substantial throughout the sub-continent \citep{bugaje_renewable_2006, diabate_solar_2004}, which explains why in rural areas, except for traditional use of solid biomass \citep{world_energy_outlook_2014}, solar energy is favored over other renewable resources, both in terms of existing facilities \citep{ren_global_status_report_2015} and promoted energy schemes \citep{bugaje_renewable_2006, chineke_political_2010}. Regarding off-grid rural electrification, renewable energy policies are also mainly relying on this potential \citep{karekezi_pv_led_renewable_2002, bugaje_renewable_2006}. Globally, electricity supply using solar energy power systems is carried out through concentrating solar power (CSP) and photovoltaic technologies; the installed PV capacity is however currently 40 times larger than the CSP capacity \citep{ren_global_status_report_2015}. In Sub-Saharan Africa, stand-alone PV systems, especially SHS, remain the most frequently encountered applications for providing electricity to remote areas as well as the most instigated in rural energy schemes \citep{ren_global_status_report_2015, karekezi_pv_led_renewable_2002, bugaje_renewable_2006, chineke_political_2010}.

\subsection{Global PV development framework: grid-connected versus off-grid systems}

Among modern renewable power systems, solar PV has benefited strongly from the new global energy policy during the last 20 years. The worldwide installed capacity has rapidly grown, from some hundred of MW in the middle of the nineties to 177\,GW by the end of 2014, standing now for the third largest renewable power generation in the world \citep{ren_global_status_report_2015}. However, this evolution hides a fundamental segmentation of the technology between both the grid-connected and off-grid architectures described beforehand.

\subsubsection{A technology market led by grid-connected systems}

As depicted previously in \figref{fig: renewable evolution}, the growth of modern renewable energies since the eighties has more particularly quickened after the establishment of the UNFCCC and the Kyoto Protocol. The latter has directly and indirectly incited industrialized countries (Annex I) to use renewable energies in order to limit their GHS emissions \citep{kyoto_protocol}. Since then, these resources have been taking advantage of the large amount of national renewable energy policies implemented thereafter to reach amounts of renewable power penetration into the energy schemes, corresponding to specific GHG emission reduction targets \citep{iea_renewable_energy_policies, ren_global_status_report_2015}. As a result, the growth of the installed PV capacity has followed the same exponential curve as the global one depicted in \figref{fig: renewable evolution}, and is actually one of the most significant among renewable energy systems \citep{ren_global_status_report_2015, eltawil_2010}. The worldwide capacity has persistently increased since the middle of the nineties, from 20\,\%/year between 1996 and 2001 to 32\,\%/year between 2001 and 2006 and 58\,\%/year between 2006 and 2011 \citep{pillot_2014, ren_global_status_report}. The global installed power has been multiplied by more than 200 since the establishment of the Kyoto Protocol, from 0.8\,GW in 1997 to 177\,GW at the end of 2014 \citep{ren_global_status_report_2015, ren_global_status_report}. However, both the energy policy framework and centralized power system structure of these developed countries have led to a specific distribution of the market between grid-connected and off-grid PV systems.

\begin{figure}[h]
\begin{center}
\includegraphics{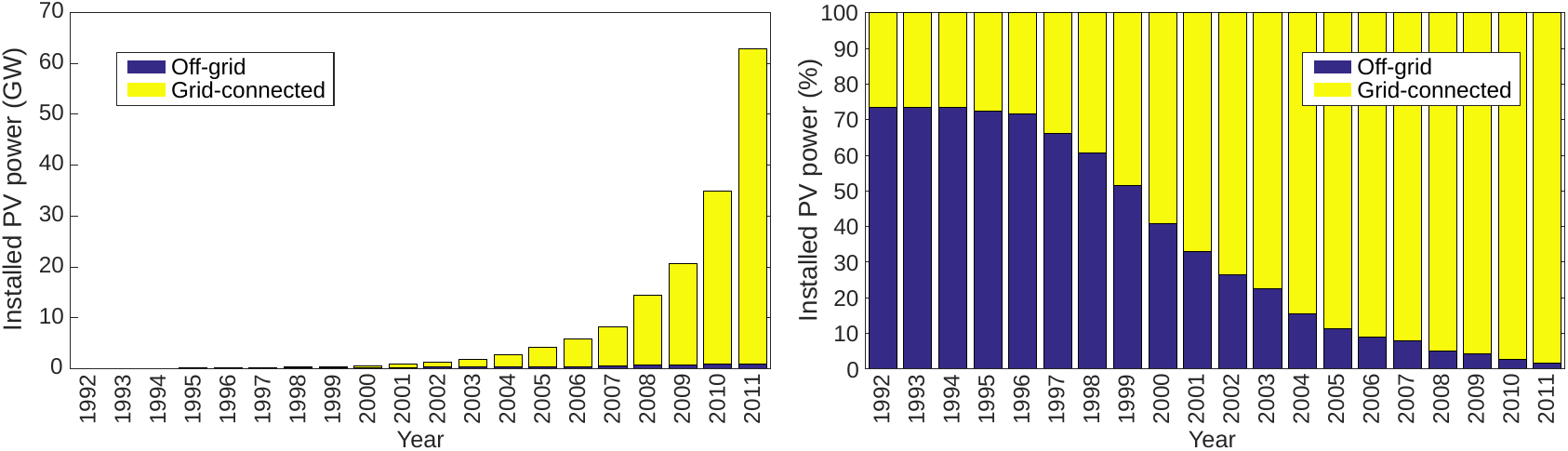}
\caption{Evolution of the off-grid/grid-connected ratio for PV systems installed in the IEA-PVPS countries between 1992 and 2011 \citep{IEA_PVPS_2012}.}
\label{fig: evolution off-grid and grid-connected PV systems}
\end{center}
\end{figure}

Historically, off-grid PV systems have been leading the sector until the mid-nineties, but still within a niche market \citep{IEA_PVPS_2012, ren_global_status_report}. At the dawn of the 21st century, many incentive measures have been initiated, such as the \emph{Renewable Energy Sources Act} in Germany \citep{jager-waldau_pv_2012, li_2015}, mainly for reaching given shares of renewable-based supply in the energy mix. Regarding PV technology, it has mainly consisted in providing a public financial support so as to balance the substantial costs of the technology and initiate the market \citep{pv_public_support_2012}. Nevertheless, these grants have primarily supplied the grid-connected systems, through attractive and guaranteed feed-in tariffs \citep{jager-waldau_pv_2014}, while stand-alone systems have been receiving very little support \citep{IEA_PVPS_2012}. This incentive policy perfectly succeeded since PV capacity has exponentially increased over the last 15 years; but this period has also stood for the success of the grid-connected over the decline of the off-grid architecture. \figref{fig: evolution off-grid and grid-connected PV systems} depicts the evolution of the PV sector according to the system configuration in the IEA-PVPS\footnote{\emph{International Energy Agency Photovoltaic Power Systems (IEA-PVPS) Programme}} countries, which currently account for more than 88\,\% of the worldwide installed PV capacity (156\,GW) against about 1\,\% in Middle-East and Africa \citep{IEA_PVPS_2015}. This figure shows that in the early 2000, when the market was emerging, the grid-connected design began to take over the off-grid model to finally reach 98.1\,\% of all the listed PV systems in 2011; the last available record (2014) gives the same ratio, with off-grid applications sharing between 0 and 2\,\% of the market \citep{IEA_PVPS_2015}. In 2011, out of the 27\,970\,MW of PV power installed in the IEA-PVPS countries, that is 93\,\% of the total yearly installed power, the share of the grid-connected systems was 99.6\,\%.

As a matter of fact, both the 100\,\% electrification rate and centralized power system architecture have clearly driven these countries to consider the easiest way for implementing renewable energies, such as PV, in their energy mix. Two main and related reasons can explain the success of the grid-connected model. First, the need for rapidly initiating and growing the market, in order to fulfill the short-term penetration level targets established by the countries' energy policies \citep{ren_global_status_report_2015, li_2015}. Second, as a way to reduce the use of costly storage systems, the possibility, in highly interconnected power networks, of balancing the inherent intermittency of renewable resources such as PV by using geographical aggregation\footnote{For example, according to Jewel \etal{} \citep{jewell_1990}, the PV penetration rate into the grid can rise from 1.3\,\% in central-station mode to a limit of 35.8\,\% when scattered throughout an area of 1000\,$\mathrm{km^2}$ or greater.}, combined with the existing use of flexible generation \citep{kaundinya_2009, li_2015, freris_renewable_2008, haurant_these_2012, eltawil_2010}.

\subsubsection{A R\emph{\&}D paradigm focusing on the grid-connected model}

Upstream and alongside the industry development, the research in PV has also grown significantly within the global energy framework. From some publications a year dealing with PV at the dawn of the fifties, this number dramatically increased after the first oil shock, to reach around 3000 publications a year between the eighties and the end of the 20th century \citep{kazmerski_1997}. For the time being, PV is the most studied technology in the field of stand-alone systems addressing rural electrification in developing countries \citep{mandelli_2016}. However, some indicators show the R\&D paragon has followed the same divarication between grid-connected and off-grid models. In fact, in the mid-nineties, Drennen \etal{} \citep{drennen_1996} considered solar PV an irrelevant economical opportunity for developing countries, and addressed need of substantial R\&D policy for including PV in the future. Since then, the technology field has dramatically grown, but the original sector economical issues don't appear to have moved on in Sub-Saharan Africa. In this way, many economic studies have been advising against using solar PV in the sub-continent until today, due to the still substantial cost of the technology \citep{karekezi_pv_led_renewable_2002, wamukonya_solar_2007, baurzhan_2016}.

Although the research generally precedes the industry, the second using commercialized version of the concepts and prototypes the first has developed, the path followed by the latter, mainly resulting from the policy framework, might also drive the R\&D paradigm. In \citep{pillot_2014}, we have analyzed the main topics of the European Photovoltaic Solar Energy Conference (EUPVSEC), which is regarded as \enquote{the largest international conference in the field of Photovoltaics} as well as a \enquote{world renowned science-to-industry platform} \citep{site_internet_eupvsec}. Storage coupled to an intermittent source is the essential element which differentiates the stand-alone from the grid-connected configuration. The second criterion is about the conditions the systems will be operating in, affecting both efficiency and reliability, such as climate, operating environment or availability of human monitoring; all or part of these conditions are different in Sub-Saharan Africa in comparison with developed countries where most of the systems are established. Analysis of the EUPVSEC emphasizes the means the research allocates to these criteria. In 2012, they accounted for approximately 8\,\% of all the oral communications \citep{EUPVSEC_programme_2012, pillot_2014}, half of them being focused on PV module degradation, the other half being dedicated to the storage (2\,\%) and to the global stand-alone system itself (2\,\%). If it confirms the shape of the PV market, it is also relevant to note that the meeting of the technology (modules + storage + electronics) with the context specific conditions (Sub-Saharan arid/tropical climates, less developed countries) appears to be a void in the research field. This pattern is also corroborated by authors having addressed either the still significant amount of research needs for meeting the off-grid system specific challenges in developing countries \citep{schafer_2011}, or the scarce existing literature about synergizing all PV facets into comprehensive off-grid application studies \citep{chaurey_2010}.

\subsection{Impact on off-grid PV sustainability in Sub-Saharan Africa}

The global energy policy over the past 40 years, coupled with the specific power supply architecture of developed countries, has led to a specific compartmentalization of the PV sector, in both industry and research, as a result that the grid-connected development is now by far prevailing over the off-grid one. Lack of knowledge about the challenges of operating stand-alone systems in a different context such as Sub-Saharan Africa has inevitably arisen in the wake of this pattern, resulting in sustainability issues for the off-grid applications effectively set up on site and uncertainty about the future of these systems.

\subsubsection{Lack of feedback and expertise}

The most obvious consequence of the historical development of the PV market around the grid-connected pillar has been the inevitable lack of feedback and expertise for stand-alone PV applications, especially in the specific context of developing countries \citep{nieuwenhout_2001, schafer_2011}. In fact, if the global part of the off-grid market is evaluated around 2\,\% \citep{ren_global_status_report}, 85\,\% of all the corresponding systems are currently located in the IEA-PVPS countries \citep{IEA_PVPS_2012}. In Africa, even if PV is promoted over other renewable resources, the sector is still in its infancy \citep{jager-waldau_pv_2014}: despite the lack of available data, the installed PV capacity, mainly consisting of small stand-alone applications, such as SHS and SPS \citep{jager-waldau_pv_2014, ren_global_status_report_2015}, was estimated under 200\,MW in 2010 \citep{epia_global_market_pv_2012, european_photovoltaic_industry_association_solar_2011}. This capacity was around 600\,MW by the end of 2013 and is increasing since then, but almost exclusively in the 2 biggest markets, that is South Africa and Algeria, and mainly supported by policies promoting utility-scale PV projects \citep{epia_global_market_pv_2015, jager-waldau_pv_2014, giglmayr_2015, ren_global_status_report_2015}.

Regarding the current operating systems, different studies have pointed out the low monitoring of rural off-grid PV projects and programs in developing countries \citep{nieuwenhout_2001, schafer_2011}. This absence of systematic evaluation of population field experience with decentralized PV systems has led to a significant void in documentary information \citep{nieuwenhout_2001, schafer_2011}. It echoes the observation made by Chaurey \etal{} on the corresponding research field \citep{chaurey_2010}: if the literature has proposed approaches for wider dissemination of off-grid PV systems, it has meanwhile poorly depicted the challenges related to marketing, dissemination and use of these systems.

The national or international level of expertise in a specific energy sector can be apprehended from its technological progress status, that is from the properly operating field applications which have been developed. For instance, Iceland is currently the world leader in geothermal installed power and production per capita \citep{ren_global_status_report_2015}, and its authority on this technology is such that other countries, like Kenya or Djibouti, draw on its expertise \citep{ahmed_aye_integration_2009, ogola_2012}. At the present time, in the rural framework, in spite of being preeminent in field projects and renewable energy policies of Sub-Saharan Africa, the PV technology remains globally confined and locally mostly present in developed countries. Also, the existing decentralized facilities seem to run anonymously, with almost no technical monitoring which would allow fast and easy feedback for enhancing system R\,\&D. This pattern coherently meets the limited investigation made by industry and research in the field of stand-alone systems, especially in the context of Sub-Saharan countries.

\subsubsection{Critical social-technological issues affecting sustainability of off-grid PV systems in the Sub-Saharan context}

At the present time, the situation we have been depicting until now is as follows: the global energy policy framework has impacted the industrial and R\&D PV sectors, leading to a tremendous underdevelopment of the PV off-grid market, especially in the Sub-Saharan context, in which there is also no transfer of experience and knowledge from the few systems operating in the field. 

Sustainability issues, and related challenges, of off-grid PV systems in Sub-Saharan Africa mainly arise at the interface of both social and technical paradigms \citep{schafer_2011}. Technical issues, deriving from the lack of industry development and system R\&D, combined with the lack of science interdisciplinary cooperation have been leading to a failure in developing the right strategies for implementing these systems in Sub-Saharan remote locations \citep{chaurey_2010, schafer_2011}.

\subsubsection*{Off-grid PV behavior in the Sub-Saharan environment}

In order to apprehend the main potential technical issues of stand-alone PV systems in Sub-Saharan Africa, we must describe the influence of exogenous parameters (operating environment) on the behavior of a PV system (efficiency, reliability, degradation). Though system interaction with wildlife, such as cables affected by mice and rats \citep{schafer_2011}, may also be accounted into operating environment, we mainly focus here on climatic conditions altering the sub-system's endogenous features. In fact, most of the climates throughout the sub-continent are either equatorial, tropical or arid \citep{liebard_climat_chaud_2005}; analysis of the behavior of off-grid PV systems operating in conditions characterized by high temperature, high humidity or else a significant amount of dust is therefore of prime importance.

Regardless of the material properties and the installation type, operating cell temperature mainly increases with incident radiation and air temperature \citep{alonso_garcia_2004}, and can significantly cut down the available power, i.e. mitigate efficiency, of a PV module \citep{gergaud_modelisation_2002, nishioka_2003}. Leaving aside the incident radiation, the PV module yield can also be mitigated by the increase of ambient air temperature alone \citep{eikelboom_2000, kim_2011}. In the same way, rise of the temperature, as well as the moisture, may strongly reduce reliability and lifetime, i.e. increase the degradation rate, of both PV fields and electric converters (charge controller, inverter) \citep{vazquez_2008, mishra_reliability_1996, sharma_2013, schafer_2011}. For the very widespread mono-crystalline and polycrystalline PV modules, the degradation rate rises for instance from a nominal value about 0.5\,\%/year \citep{jordan_2013, osterwald_2006} to 1\,\%/year in arid climates \citep{bogdanski_2010, jordan_2013}, and appears even more substantial in Sub-Saharan environmental conditions, up to 3\,\% \citep{ndiaye_2014}. In relation to converters, as charge regime is a critical point for maintaining battery service life \citep{al-sheikh_2012}, malfunctioning charge controllers have also been found to be often responsible for failure and lifetime mitigation of batteries \citep{nieuwenhout_2001}.

Dust's major contribution is related to the PV module efficiency decrease due to soiling \citep{sayyah_2014, ndiaye_2013b, kazem_2014}, which has led several authors to address requirements for PV module cleaning, in particular integrated into plans of periodic preventive maintenance \citep{ndiaye_2013b, sayyah_2014}. Dust's minor contribution concerns its impact on PV panel degradation over lifetime, which has been for example estimated around 16-29\,\% of the global power losses by Tanesab \etal{} \citep{tanesab_2015}.

Like the other components of the system, performances of the electrochemical storage also strongly depends on these criteria. With reference to aging, battery nominal lifetime is thus halved for each increase of $10\,\degre\mathrm{C}$ above the reference operating temperature \citep{magnor_concept_2009, svoboda_2007}, and rainy seasons of equatorial and tropical countries are also a source of damage, and then of lifetime mitigation \citep{schafer_2011}. Especially, in addition to this statement, the very used lead-acid batteries already present short nominal lifetimes combined with intolerance to extreme temperatures \citep{daim_2012}. This is a cause for concern as it makes the batteries still standing for the major contribution to the life-cycle investment of a stand-alone PV system \citep{hegedus_status_2003, nassen_2002, nieuwenhout_2001}; their multiple replacement along the overall system lifetime reduces its economic relevance.

\subsubsection*{Socioeconomic implications of the stand-alone PV system R\emph{\&}D requirements in Sub-Saharan Africa}

The main socioeconomic implications spring out from the light thrown on the stand-alone PV system reliability, degradation and performance issues within the specific Sub-Saharan rural context, deriving from the impact of the operating environment on the material endogenous features. The extreme remoteness of the populations makes these technical issues especially critical as maintaining performances and reliability requires in particular monitoring systems, performing repair and maintenance or else providing spare parts \citep{schafer_2011, nieuwenhout_2001, chaurey_2010}. The corresponding challenges reflect the lack of PV transdisciplinary R\&D \citep{schafer_2011}. Regarding system promotion, installation and maintenance, lack of skilled technicians or knowledge transfer between PV producers and fitters/users is pointed out; system failures thereby often result from unsuitable design, improper installation, use of unreliable components or poor maintenance \citep{schafer_2011, jafar_2000, chaurey_2010, mulugetta_2000}. Besides, users are also often left on their own with their system after the initial installation, without any real after-sales service structure or proper training \citep{baurzhan_2016}. Technology-user interaction is therefore addressed as a critical point in the literature \citep{chaurey_2010, schafer_2011, jafar_2000, mulugetta_2000}: maintaining system proper performances and reliability in remote areas requires user operation training and significant understanding of the very specific characteristics of the system, but still appears to be left out. Daily deteriorated system performances resulting from operating environment, especially owing to the batteries \citep{gustavsson_2004}, are another cause for concern. Disappointment may for example arise from the differential between user expectations and the system energy performance ultimately delivered on site, as the installation cannot always supply the intended user needs because of the incomplete or wrong information about its real-conditions operating capacities \citep{schafer_2011, chaurey_2010}.

Another consequence is the very high cost of off-grid PV systems in Sub-Saharan Africa, which affects related financing schemes used for system dissemination, mainly due to operating and maintenance charges required for counterbalancing the aforementioned reliability and performance issues \citep{chaurey_2010, baurzhan_2016, wamukonya_solar_2007}. Potential users are often unaware of these extra costs in addition to the PV system initial investment, as they are generally underestimated in financial schemes, and have therefore to cope with it while they are repaying the credit \citep{baurzhan_2016}. For instance, specific electrochemical storage short-term issues can lead stand-alone systems to operate with poor batteries \citep{gustavsson_2005}, or for the users to be forced to give up on their installation after battery failure and revert to their previous energy supply while still paying system investment credits \citep{schafer_2011}. As a result, the first targeted households for using these systems are also the ones that can hardly afford the cost of such facilities \citep{baurzhan_2016}, which therefore require innovative financing schemes such as fee-for-service or donation delivery mechanism \citep{lemaire_2009, chaurey_2010, nieuwenhout_2001}. However, when off-grid PV projects rely on the latter, they generally fail because of lack of user commitment, as they are less involved and feel less responsible than for a purchased system \citep{nieuwenhout_2001}: maintenance for sustainable operation is not originally understood, savings for battery replacement are not necessarily made, and use of the system is therefore more easily abandoned.

\section{Conclusion and prospects}

The \emph{sustainable development} concept sprung up in the developed countries, in the wake of the questioning, which emerged in the seventies, on the impact of their historical development on the living environment \citep{meadows_1972, brundtland_1987}. It has been shaping the global policy framework since the nineties, starting with establishment and ratification of both UNFCCC and Kyoto Protocol by most of the countries in the world \citep{unfccc_1992, kyoto_protocol}. This global framework has been then implemented locally in national climate change policies, from developed to developing countries, including Sub-Saharan Africa \citep{rapport_ministere_djibouti_ccnucc_2001, bugaje_renewable_2006, chineke_political_2010, reid_1998}. In this way, GHG emission reduction targets of the Kyoto Protocol have directly and indirectly supported the use of renewable power sources, resulting in many local policies having sparkled the tremendous growth of these resources, especially the modern sector such as wind or solar PV \citep{iea_renewable_energy_policies, ren_global_status_report_2015, IEA_energy_balance_non_OCDE_2014}. However, both the power system centralized architecture and the 100\,\% electrification rate of the most developed regions in the world, directly stated by the Kyoto Protocol disposals for reducing their GHG emissions, has led to a significant divarication of the renewable power supply global market between grid-connected and off-grid sectors. This is particularly manifest for solar PV systems, which have moved on from a narrow market dominated by off-grid facilities in the mid-nineties to the third renewable power installed capacity in the world in 2014, mostly in developed countries, with the grid-connected design sharing for more than 98\,\% of all installations \citep{IEA_PVPS_2012, ren_global_status_report_2015}. In short, the need of developed countries for reaching short-term GHG emission reduction targets, according to \emph{sustainable development} requirements, has driven industry and R\&D to focus on the grid-connected architecture. This model allows replacement of traditional fossil fuel plants by renewable power systems in highly interconnected electrical grids, taking advantage of geographical aggregation and existing flexible generation instead of costly storage systems in order to limit intermittency issues \citep{kaundinya_2009, li_2015, freris_renewable_2008, haurant_these_2012, eltawil_2010}. On the other hand, the need of Sub-Saharan Africa for increasing rural access to electricity, in order to enhance \emph{human development} of most of the population, has led to the consideration of off-grid electrification \citep{adkins_2010, pillot_2014, lahimer_2013, mandelli_2016, huacuz_photovoltaics_2003, pillot_ecm_2013}.

This global framework has shaped the current status of off-grid renewable power systems in Sub-Saharan Africa. Regarding solar PV, although it remains the most encountered technology in the field and the most instigated in both R\&D and energy incentive policies \citep{mandelli_2016, ren_global_status_report_2015, bugaje_renewable_2006, chineke_political_2010}, the installed capacity is still very low and the last improvements have mainly affected utility-scale facilities \citep{epia_global_market_pv_2015, jager-waldau_pv_2014, giglmayr_2015, ren_global_status_report_2015}. Feedback about how these few systems are in fact operating in the specific Sub-Saharan field has been definitely lacking , and research has meanwhile been following the industrial path by leaving aside the off-grid system specificities. As a result, R\&D original challenges and needs, arising at the interface of both technological and socioeconomic issues in the Sub-Saharan context, appear to have been left untouched \citep{schafer_2011, nieuwenhout_2001, chaurey_2010}.

About 95\,\% of the renewable energy policies have followed the UNFCCC and 90\,\% the Kyoto Protocol \citep{iea_renewable_energy_policies}; it makes therefore really questionable the original goal behind the implementation of these resources in Sub-Saharan countries' energy schemes. In the region, it seems indeed it has never been really clear whether renewable energies were an artifact from and for sustainable development and climate change policies or an electrification mean to address poverty mitigation \citep{rapport_ministere_djibouti_ccnucc_2001, bugaje_renewable_2006, chineke_political_2010, reid_1998, karekezi_renewables_meeting_poors_2002, baurzhan_2016}. This somehow \emph{schizophrenia} is especially apparent when confronting the aforementioned promotion and use of solar PV for electrifying remote areas of the sub-continent with the literature having meanwhile considered it economically unfeasible over the last 20 years \citep{drennen_1996, baurzhan_2016, karekezi_pv_led_renewable_2002, wamukonya_solar_2007, mulugetta_2000}. The paradox arising from this situation is two-fold, so that neither sustainable development nor human development requirements have been really fulfilled in Sub-Saharan Africa. As a matter of fact, in the region, so-called \emph{sustainable} off-grid energy power systems remain scarce and subject to sustainability failures\footnote{Note that this is actually also somehow true for most of the grid-connected systems at the moment, as power networks and infrastructures remain particularly unreliable in Sub-Saharan Africa, and therefore technically unsuited for merging substantial intermittent renewable power \citep{moyo_2013, karekezi_renewables_meeting_poors_2002}.}, while multidimensional poverty is still as significant as the electrification rate is low \citep{HDR_2014, world_energy_outlook_2014}.

The success, or sustainability, of renewable power systems in either case most probably lies in focusing on how to reach social, economic and political aims given the specific and relevant features of the context. In other words, renewable sources cannot be a \emph{de facto} solution, merely based on political will; the choice of whether to implement them or not must instead rely on any kind of fair cost-benefit analysis\footnote{It is important to note that \emph{cost-benefit} not only refers to economics but also to any relevant field, such as social, energy, environment, etc.}. It has been especially true, for instance, for sustainable development in the developed countries, as those resources have been standing for a means towards GHG emission reduction targets and exploited accordingly. Likewise, the best path towards success in sustaining renewable systems in Sub-Saharan Africa shall presumably be to focus on the primary objective to be achieved: \emph{durably enhancing human development of the corresponding populations}. Subsequently, regarding rural off-grid electrification, pros and cons of all alternatives, among which renewable energies, would be reviewed so as to get and implement the optimal\footnote{Here, we refer to minimizing the theoretical cost-benefit function associated with the objective of long-term multidimensional poverty mitigation.} option, or set of options, at the moment, as well as to improve the others in the long run.

As an opening to this global overview, we hence might ask: should human and sustainable developments be \emph{decoupled} in Sub-Saharan Africa, such that renewable resources would be just one of many agents in achieving rural off-grid electrification, or can we make both, and their corresponding requirements, more compatibles ?

The first part of this question requires us to put the global sustainability issue into perspective. Between 1850 and 2005, the most developed countries (HDI > 0.8) have been responsible for almost 64\,\% of all CO$_2$ emissions \citep{HDR_2011, pillot_2014}. For now, the gap in GHG emissions, primary energy supply or electricity consumption between Sub-Saharan Africa and developed countries is still tremendous; as an example, IEA countries, having originally supported the current global energy framework, stand for amounts respectively 7, 14 and 21 times larger than Sub-Saharan countries \citep{pillot_2014}. Not only the historical responsibility in both global warming and resources depletion issues but also the current situation are therefore completely unbalanced between Sub-Saharan and developed regions. Actually, the significant human development of the most industrialized regions in the world has given them the latitude to rethink their development path, whose historical global consequences seem to have restrained the latitude of the developing regions to achieve their own.

From this pattern, it is probably legitimate to consider both the paragons having to get along with each other, using their own \emph{fitted means} to limit, for developed regions, negatively affecting their environment (sustainable development) and increase, for Sub-Saharan Africa, the population's standard of living (human development). Nevertheless, as a first step in answering the second part of the original question, we may remind ourselves what we depicted in section \ref{sec: electricity and qualitative part of human development}: GHG emissions and primary energy supply are mainly related to economic growth, while enhancing education and health criteria of human development is of lower influence. In the first stages of diminishing multidimensional poverty, by removing deprivations in every dimension accordingly, electrification would therefore be a \emph{qualitative} energy supply vector whose environmental impact would remain circumscribed. This could be then the first step towards aware populations, equipped with the right instruments for promoting a different path, a \emph{sustainable} path, towards the next stage of their development.

% suited means, adjusted means, etc.

 %Typically, achieving sustainable development in fully electrified regions using centralized power networks has led to considering grid-connected systems.  

% In other words, how to promote the outs of renewable energies, that is the possible positive influence on human development through rural off-grid electrification, and not themselves, as a political instrument imported into a different local context ?

\section{References}
\bibliographystyle{elsarticle-num}
\bibliography{References}

\end{document}